\documentclass[twocolumn,twoside,slac_two]{revtex4}
\usepackage{graphicx}
\usepackage{fancyhdr}
\def \Bbar{\bar B}
\def \fbar{\bar{f}}
\def \Kbar{\bar K}
\def\bd{B_d^0}
\def\bs{B_s^0}
\def\sss{\scriptscriptstyle}

\def\ApNPqph{{\cal A}^{\prime,q} e^{i \Phi'_q}}
\def\ApNPCqph{{\cal A}^{\prime {\sss C}, q} e^{i \Phi_q^{\prime C}}}
\def\ApNPCuph{{\cal A}^{\prime {\sss C}, u} e^{i \Phi_u^{\prime C}}}
\def\ApNPCdph{{\cal A}^{\prime {\sss C}, d} e^{i \Phi_d^{\prime C}}}
\def\pewcp{P_{\sss EW}^{\prime\sss C}}
\def\pewp{P'_{\sss EW}}
\def\ApNPuph{{\cal A}^{\prime,u} e^{i \Phi'_u}}
\def\ApNPdph{{\cal A}^{\prime,d} e^{i \Phi'_d}}
\def\ApNPcomb{{\cal A}^{\prime, comb} e^{i \Phi'}}

\begin{document}

%Title of paper
\title{U-spin Implication for $B_s$ Physics and New Physics}

% Repeat the \author .. \affiliation  etc. as needed
%
% \affiliation command applies to all authors since the last
% \affiliation command. The \affiliation command should follow the
% other information

\author{Makiko Nagashima}
\affiliation{
Physique des Particules, Universit\'e de
Montr\'eal,C.P. 6128, succ. centre-ville, Montr\'eal, QC,
Canada H3C 3J7}

\begin{abstract}
With U-spin symmetry, $b\to s$ and $b\to d$ penguin decays could be a subtle
probe of CP violating new physics contributions.
We show that, for $B\to PP$ ($P$ stands for a pseudoscalar meson),
the U-spin relation is expected to be violated for only one decay pair
by assuming that new physics affects only $b\to s$ transition processes.
We also very shortly discuss the polarizations of two types of U-spin
pairs for $B\to VV$ ($V$ stands for a vector meson).
\\
{\it This article describes my poster presentation at the $6^{\rm th}$ 
Flavor Physics and CP Violation Conference held at National Taiwan University
(5/5/08 $\sim$ 5/9/08).}
\end{abstract}

%\maketitle must follow title, authors, abstract
\maketitle

\thispagestyle{fancy}

% body of paper here - Use proper section commands
% References should be done using the \cite, \ref, and \label commands
% Put \label in argument of \section for cross-referencing
%\section{\label{}}

\section{Introduction}
% 1st para
Within the past decade, several different experiments have been focusing
on the measurements of CP violation in various flavor violating $B$ decays.
The main aim is to test the Standard Model (SM) which contains only one source
of CP violation.
% in the weak vertex.
A definitive discrepancy between experimentally measured and theoretically
predicted CP asymmetries in $B$ decays, if found, would indicate the presence
of new CP violating sources beyond the SM.

% 2nd para
According to a recent UT$fit$ study about the $B_s$ mixing phase \cite{UTfit},
together with updated data from Belle for $B_{u,d}\to \pi K$
\cite{BelleNature},
$b\to s$ transition is seemingly affected by
CP violating new physics (NP). However, none of large discrepancies
with the SM has been found in $b\to d$ phenomena at present \cite{hfag}.
In fact, the CP violation in $B_s\to \pi K$, recently measured by CDF,
would agree with the SM expectation \cite{CDFreport}.

% 3rd para
It is worth mentioning that the strong hint of NP is indicated in certain
$b\to s$ CP violating phenomena.
If the current observations are confirmed by further experimental
improvements, what kind of NP scenario appears to be plausible?
Throughout our presentation, we assume that U-spin symmetry reasonably holds
among different B decays. We then point out that the similar hint of NP can 
be found by comparing $B_d\to K^0\pi^0$ with $B_s\to\bar K^0\pi^0$.
We also very shortly discuss the polarizations of $B_s\to K^{*0}\bar K^{*0}$,
$\phi \bar K^{*0}$.

% 4th para % paper is organized as follows:

\section{U-spin symmetry and Standard Model \label{sec:secII}}
% 1st para
Sometime ago it was pointed out that, within the SM, a relation involving
decay rates and direct CP asymmetries holds for the $B$ decay pairs that
are related by U-spin \cite{Fleischer99, Gronau00}.
U-spin is the symmetry that places $d$ and $s$ quarks on an equal footing.
The pairs associated with $B\to \pi K$ are listed in Table~\ref{tab:UspinPair},
as well as two pairs of neutral $B$ decays into two vector mesons,
which are discussed in our presentation.

\begin{table}[t]
\center
\begin{tabular}{cccc}
\hline\hline
 & $\bar b\to \bar s$ & $\longleftrightarrow$ & 
$\bar b\to \bar d$ \\ \hline
%PP& & & \\ 
\#1 & $\bd \to K^+\pi^-$ & $\longleftrightarrow$ & $\bs \to \pi^+ K^-$ \\
%\#2 & $\bs \to K^+ K^-$& $\longleftrightarrow$ & $\bd \to \pi^+\pi^-$ \\
\#2 & $\bd \to K^0\pi^0$ & $\longleftrightarrow$ & $\bs \to \Kbar^0\pi^0$ \\
\#3 & $B^+ \to K^0 \pi^+$ & $\longleftrightarrow$ & $B^+ \to \Kbar^0 K^+$ \\
%\#5 & $\bs \to K^0 \Kbar^0$ & $\longleftrightarrow$ & $\bd \to \Kbar^0 K^0$ \\
%\#6 & $\bs \to\pi^+\pi^-$ & $\longleftrightarrow$ & $\bd \to K^+ K^-$ \\
%\hline
%VV& & & \\
\#4 & $\bs \to K^{*0} {\bar K}^{*0}$ & $\longleftrightarrow$ & 
$\bd \to {\bar K}^{*0} K^{*0}$ \\
\#5 & $\bd \to\phi K^{*0}$ & $\longleftrightarrow$ & 
$\bs \to\phi {\bar K}^{*0}$ \\ \hline\hline
\end{tabular}
\caption{The pairs of $B$ decays which are related by U-spin.
\#1-\#3 are for $B$ decays into two pseudoscalar mesons \cite{Gronau00},
while \#4-\#5 are for $B$ decays into two vector mesons.}
\label{tab:UspinPair}
\end{table}
%

% 2nd para
In the limit of U-spin symmetry, the effective Hamiltonian
describing a $b\to d$ transition is equal to that of the corresponding
$b\to s$ transition, where the elements of the Cabibbo-Kobayashi-Maskawa
(CKM) matrix are changed appropriately. 
With the CKM unitarity relation \cite{Jarl},
\begin{eqnarray}
{\rm Im}(V^*_{ub}V_{us}V_{cb}V^*_{cs}) = 
- {\rm Im}(V^*_{ub}V_{ud}V_{cb}V^*_{cd}),
\end{eqnarray}
a perfect U-spin symmetry guarantees the following relation,
\begin{eqnarray}
& &\vert A(B \to f)\vert^2 - \vert A(\Bbar \to \fbar)\vert^2 
\nonumber \\
&=& 
 -\left[ \vert A(UB \to Uf)\vert ^2 - \vert A(U\Bbar \to U\fbar) 
\vert^2 \right],
\label{eq:Urel}
\end{eqnarray}
in which $U$ is the U-spin operator that transposes $d$ and $s$ quarks.
This expression can be written as
\begin{eqnarray}
{- A_{\sss CP}^{dir}{\hbox{(decay \#1)}} \over A_{\sss CP}^{dir}
{\hbox{(decay \#2)}}} =
{BR{\hbox{(decay \#2)}} \over BR{\hbox{(decay \#1)}}},
\label{eq:Urelmod} 
\end{eqnarray}
where $A_{\sss CP}^{dir}$ and $BR$ refer to the direct CP asymmetry
and branching ratio, respectively, and where decays \#1,2 are the
$b\to d$ and $b\to s$ decays, in either order, related by U-spin. 
Note that if decays \#1,2 include $B_d$ and $B_s$ mesons,
there is an additional factor on the right-hand side taking the
lifetime difference into account.

% 3rd para
Let us now look at pair \#1. Ref.~\cite{Lipkin05} has paid special
attention to this pair and argued about a test of the SM vs. NP.

% 4th para
The decay amplitudes for pair \#1 are given by
\begin{eqnarray}
& & A(B_d\to K^+\pi^-)= V^*_{ub}V_{us}A_u^{(s)} + V_{cb}V^*_{cs}A_c^{(s)},
\nonumber \\
& & A(B_s\to \pi^+K^-)= V^*_{ub}V_{ud}A_u^{(d)} + V_{cb}V^*_{cd}A_c^{(d)},
\end{eqnarray}
where $A_{u}^{(q)}$ and $A_{c}^{(q)}$, with $q=s,d$, are strong decay factors
from tree and penguin amplitudes. The index $q$ refers to the $b\to q$
transition in the penguin diagrams.
To date there is no firm evidence, in $B$ decays into two
pseudoscalar mesons, that nonfactorizable contributions (including penguin
annihilation effects \cite{BN2003}) are sizable.
We therefore employ a picture where the nonfactorizable terms are simply
higher-order contributions and less important than the (naive) factorizable
terms.
As long as this picture reasonably holds, the U-spin breaking effects
are expected to be small.
One then finds that $A_u^{(s)}=A_u^{(d)}=A_u$ and $A_c^{(s)}=A_c^{(d)}=A_c$ 
are good approximations, and Eq.(\ref{eq:Urelmod}) is satisfied for pair \#1
within the SM. In the following discussion,
we basically follow the same strategy.

\begin{table}[t]
\center
\begin{tabular}{lcc}
\hline
\hline
Mode &  $BR[10^{-6}]$ & $A_{\sss CP}$[$\%$] \\ \hline
$B^+ \to \pi^+ K^0$ & $23.1 \pm 1.0$ & $0.9 \pm 2.5$  \\
$B^+ \to \pi^0 K^+$ & $12.9 \pm 0.6$ & $5.0 \pm 2.5$  \\
$\bd \to \pi^- K^+$ & $19.4 \pm 0.6$ & $-9.7 \pm 1.2$  \\
$\bd \to \pi^0 K^0$ & $9.9 \pm 0.6$ & $-14 \pm 11$  \\
$B_s \to \pi^+ K^-$ & $5.00 \pm 1.25$ & $39 \pm 17$  \\
\hline
\hline
\end{tabular}
\caption{Branching ratios and direct CP asymmetries for the
$B\to \pi K$ decays. The data is taken from Ref.~\cite{hfag}.}
\label{tab:expdataKpi}
\end{table}

% 5th para
Recent experimental results for $B\to \pi K$ are collected in
Table~\ref{tab:expdataKpi}. Using these data, we can estimate
the U-spin relation for pair \#1, and it turns out
\begin{eqnarray}
{- A_{\sss CP}^{dir}(\bs \to \pi^+ K^-) \over
   A_{\sss CP}^{dir}(\bd \to \pi^- K^+)}  &=&  4.2 \pm 2.0 ~~,
\nonumber \\
{BR(\bd \to \pi^- K^+) \over BR(\bs \to\pi^+ K^-)}  &=&  3.9 \pm 1.0. 
\label{eq:pairone}
\end{eqnarray}
Although the error is still large, the two ratios are nearly equal.
It is remarkable that the Eq.~(\ref{eq:Urelmod}) is satisfied for pair \#1,
{\it i.e.} there is apparently no evidence of NP in this pair.
Is the fact found in pair \#1 contradictory to other clues observed in
different $b\to s$ phenomena?
We would like to emphasize that this could be another interesting hint for NP.

\section{Implication for $B_s$ Physics \label{sec:secIII}}
% 1st para 
As we mentioned, to date there has been no visible discrepancies with the SM
in $b\to d$ decays. We follow this experimental indication and assume
that the NP appears only in $b\to s$ decays but does not affect $b\to d$
decays.

% 2nd para
There are many NP operators which can contribute to $b\to s$ decays.
However, it was recently shown in Ref.~\cite{DL:NPope04} that this number
can be reduced considerably.
At the quark level, each NP contribution to the decay $B\to f$ takes the
form $\langle f| {\cal O}_{\sss NP}^{ij,q}|B\rangle$, where
${\cal O}_{\sss NP}^{ij,q} \sim {\bar s} \Gamma_i b \, {\bar q}
\Gamma_j q$ ($q = u,d,s,c$), in which the $\Gamma_{i,j}$ represent
Lorentz structures, and color indices are suppressed.
Each NP matrix element can have its own weak and strong phase.
Now, it should be mentioned that the idea of small NP strong phases
can be justified \cite{DL:NPope04,DLmethods,NSL:Uspin}.
We then further simplify our study by neglecting all NP strong phases.
One can then combine all NP matrix elements into
a single NP amplitude, with a single weak phase:
\begin{eqnarray}
\sum \langle f| {\cal O}_{\sss NP}^{ij,q} |B\rangle = 
{\cal A}^q e^{i \Phi_q} ~.
\end{eqnarray}

% 3rd para
In fact, for $b\to s$ decays, there are two classes of NP
operators, differing in their color structure: ${\bar s}_\alpha
\Gamma_i b_\alpha \, {\bar q}_\beta \Gamma_j q_\beta$ and ${\bar
s}_\alpha \Gamma_i b_\beta \, {\bar q}_\beta \Gamma_j q_\alpha$. The
first class of NP operators contributes with no color suppression to
final states containing ${\bar q}q$ mesons.
Similarly, for final states with ${\bar s} q$ mesons,
the roles of the two classes of operators are reversed, but there is a
suppression factor of $1/N_c$.
As in Ref.~\cite{DLmethods}, we denote by $\ApNPqph$ and
$\ApNPCqph$ the sum of NP operators which contribute to final states
involving ${\bar q}q$ and ${\bar s}q$ mesons, respectively (the primes
indicate a $b\to s$ transition).
Here, $\Phi'_q$ and $\Phi_q^{\prime {\sss C}}$ stand for the NP weak phases.

% 4th para
Let us now return to $B_{u,d}\to \pi K$ decays in order to study the
effect of the NP operators on $b\to s$ transitions.
In Ref.~\cite{GHLR}, the relative sizes of the SM $B_{u,d}\to \pi K$ 
diagrams were roughly estimated as
\begin{eqnarray}
& & 1 : |P'_{(c)}|,~~~~ {\cal O}({\bar\lambda}) : |T'|,~|\pewp|,~~~~
\nonumber \\
& & \;\;\;\;\;
{\cal O}({\bar\lambda}^2) : |C'|,~|P'_{(u)}|,~|\pewcp|,
\label{eq:theoHier}
\end{eqnarray}
where ${\bar\lambda} \sim 0.2$. The diagrams $T'$ and $C'$ are
the color-favored and color-suppressed trees, $P'_{(c)}$ and $P'_{(u)}$
are the gluonic penguins, and $\pewp$ and $\pewcp$ are the color-favored
and color-suppressed electroweak penguins, respectively.
Especially for the ratio $|C'/T'|$,
the SM predictions from different calculation approaches
would agree to the naive estimation of Eq.~(\ref{eq:theoHier})
\cite{BN2003,PQCD:NLO,SCET}.

% 5th para
Putting all diagrams together, Ref.~\cite{BaekLondon} had performed
a fit by using the 2006 $B\to \pi K$ data \cite{hfag}. A good fit is found,
however, this fit requires $|C'/T'|\sim 1.6\pm 0.3$ which is much larger
than the estimates of Eq.(\ref{eq:theoHier}).
This might imply that the $B\to \pi K$ fit including NP amplitudes
is necessary. If one ignores the small ${\cal O}({\bar\lambda}^2)$
diagrams, the $B\to \pi^i K^j$ amplitudes ($i,j$ are electric charges)
can be written \cite{DLmethods}
\begin{eqnarray}
A^{+0} &\!\!=\!\!& -P'_{(c)} \!+\! \ApNPCdph  \nonumber\\
\sqrt{2} A^{0+} &\!\!=\!\!& P'_{(c)} \!-\! T' \, e^{i\gamma} \!-\!
\pewp \!+\! \ApNPcomb \!-\! \ApNPCuph  \nonumber \\
A^{-+} &\!\!=\!\!& P'_{(c)} \!-\! T' \, e^{i\gamma} \!-\! \ApNPCuph
\nonumber \\
\sqrt{2} A^{00} &\!\!=\!\!& -P'_{(c)} \!-\! \pewp
\!+\! \ApNPcomb \!+\! \ApNPCdph, 
\end{eqnarray}
where $\gamma$ is the SM weak phase and
$\ApNPcomb \equiv - \ApNPuph + \ApNPdph$.
It is not possible to distinguish the two component
amplitudes in $B\to \pi K$ decays. We therefore denote all possible
NP amplitudes in $B\to \pi K$ as $\ApNPCuph$, $\ApNPCdph$, and $\ApNPcomb$.

% 6th para
The three NP operators were then included in the $B\to \pi K$ fit in
Ref.~\cite{BaekLondon}, one at a time. It was found that the fit remained
poor if $\ApNPCuph$ or $\ApNPCdph$ was added.
That is, these NP operators can be large or small. If we neglect
$\ApNPCuph$ and $\ApNPCdph$ by assuming that they are small,
a good fit was obtained through a large value of $\ApNPcomb$.
It is worth mentioning that, if we look at the amplitude $A^{-+}$ describing
$\bd \to \pi^- K^+$, the amplitude does not contain NP effects
because of $\ApNPCuph=0$. As be seen in Eq.~(\ref{eq:pairone}),
the current data is consistent with the SM prediction.
It seemingly supports the idea that $\ApNPCuph$ is small.
Therefore, we expect that $\ApNPcomb$ brings sizable effects into the $b\to s$
transition, while ${\cal A}^{\prime {\sss C}, u(d)}$ is less important.

% 7th para
In this case, it turns out that only the decay pair \#2 can be
significantly affected by NP. We would like to repeat that
this consequence is consistent with
what we have found in Sec.~\ref{sec:secII}.
The pair \#2, therefore, should be looked at more closely although the
precise measurements of time-dependent CP asymmetries for those decays
would be challenging.
%(See Ref.~\cite{NSL:Uspin} for more details.)

% 8th para
Before closing this section, we briefly refer to the U-spin implications
for $B_s\to K^{*0}\bar K^{*0}$ and $B_s\to \phi \bar K^{*0}$.
The $B_d\to \phi K^{*0}$ had brought forth the polarization puzzle
- the longitudinal and transverse components are roughly equal size -,
which cannot be explained by the {\it naive} factorization calculation
within the SM ($n$SM). There are a lot of studies dealing with this puzzle
\cite{QCDf:VV}.
Let us now assume that the transverse amplitude is expressed as a
{\it single dominant} contribution which arises from beyond the $n$SM,
and it is nonfactorizable.
As long as U-spin symmetry reasonably holds among the strong decay factors
of the final state $VV$, the transverse components in pairs \#4 and \#5
can be simply given by
\begin{eqnarray}
\frac{{\cal A}_{\sss T}(\bs \to K^{*0} {\bar K}^{*0})}
  {{\cal A}_{\sss T}(\bd \to {\bar K}^{*0} K^{*0}
  )}
& \approx &
\frac
{\left\vert V_{ts} \right\vert}
{\left\vert V_{td} \right\vert}
\frac{f_{\sss \bs}}{f_{\sss \bd}},
\nonumber \\
\frac{{\cal A}_{\sss T}(\bd\to \phi K^{0*})} {{\cal A}_{\sss
    T}(\bs\to \phi {\bar K}^{0*})} 
& \approx &
\frac
{\left\vert V_{ts} \right\vert}
{\left\vert V_{td} \right\vert}
\frac{f_{\sss\bd}}{f_{\sss\bs}}.
\end{eqnarray}
Because the final states are self-conjugate,
pairs \#4 and \#5 are expected to provide better results in terms of
theoretical uncertainties.
Consequently, one has
\begin{eqnarray}
{f_{\small T}(\bs \to K^{*0} {\bar K}^{*0}) \over 
f_{\small T}(\bd \to {\bar K}^{*0} K^{*0})} & \approx &
28.4 \pm 7.2 \, {BR(\bd \to {\bar
K}^{*0} K^{*0}) \over BR(\bs \to K^{*0} {\bar K}^{*0})},
\nonumber\\
& & 
\nonumber\\
{f_{\small T}(\bd \to \phi K^{*0}) \over 
f_{\small T}(\bs \to \phi {\bar K}^{*0})} & \approx & 
18.8 \pm 4.8 \, {BR(\bs \to \phi {\bar
    K}^{*0}) \over BR(\bd \to \phi K^{*0})}.
\nonumber\\
\end{eqnarray}
However, they might not be robust estimates since U-spin breaking in
nonfactorizable contributions would be more complicated and it might be large
\cite{footnote}.

\section{Summary \label{sec:summary}}
% 1st para
There are recent intriguing studies regarding whether a strong hint that
implies NP in $b\to s$ transition processes would be observed.
With U-spin symmetry, we speculated on the implications for $B_s$ decays.

% 2nd para 
Taking the fit results in Ref.~\cite{BaekLondon} into account,
only one NP amplitude is found to be large.
We have shown that, Eq.(\ref{eq:Urelmod}) is expected to be violated
by only one decay pair: $B_d^0\to K^0\pi^0$ and $B_s^0\to \bar K^0\pi^0$.

% 3rd para
We also shortly referred to the U-spin implications for the polarizations of
$B_s\to K^{*0}\bar K^{*0}$, $\phi \bar K^{*0}$. In the limit of
U-spin symmetry, assuming the particular scenario,
the transverse components for $B_s\to K^{*0}\bar K^{*0}$, $\phi \bar K^{*0}$
can be estimated with relatively less uncertainties.

% If you have acknowledgments, this puts in the proper section head.
\bigskip % extra skip inserted
\begin{acknowledgments}
\end{acknowledgments}
This article is based on the works in collaboration with
A.~Datta, D.~London, J.~Matias, and A.~Szynkman.
I would like to thank the conference organizers
for an excellent conference.

\bigskip % extra skip inserted
% Create the reference section using BibTeX:
%\bibliography{basename of .bib file}
%\begin{thebibliography}{9}   % Use for  1-9  references

\end{document}